\documentclass[12pt,a4paper]{article}
\usepackage{cite,mathtools,amsmath,amssymb,graphicx,subfigure}
\usepackage[usenames]{color}
\usepackage[bookmarksopen,colorlinks=true,linkcolor=dark_green,citecolor=dark_red,urlcolor=dark_red,linktocpage=false]{hyperref}

\definecolor{dark_blue}{rgb}{0,0,0.6}
\definecolor{dark_green}{rgb}{0,0.4,0}
\definecolor{dark_red}{rgb}{0.6,0,0}

\usepackage[height=22.5cm,width=16.5cm,centering]{geometry}

\def\thefootnote{\fnsymbol{footnote}}
\newcommand{\vect}[1]{\mbox{\boldmath${#1}$}}

\newcommand{\beq}{\begin{align}}
\newcommand{\eeq}{\end{align}}
\newcommand{\beqa}{\begin{eqnarray}}
\newcommand{\eeqa}{\end{eqnarray}}
\newcommand{\mpl}{M_{\rm pl}}
\newcommand{\mtilde}{\tilde{M}}
\newcommand{\lmk}{\left(}
\newcommand{\rmk}{\right)}

\newcommand{\cH}{{\mathcal H}}

\begin{document}

\begin{titlepage}

\begin{center}
\hfill CTPU-PTC-18-43 \\
\hfill LDU-18-007 \\
\hfill RESCEU-16/18 \\

\vskip 12mm

{\fontsize{20pt}{0pt} \bf
On the violent preheating
}
\\[3mm]
{\fontsize{20pt}{0pt} \bf
in the mixed Higgs-$R^2$ inflationary model
}

\vskip 10mm

{\large
Minxi He,\footnote{
Email: {\tt hemxzero"at"resceu.s.u-tokyo.ac.jp}
}$^{1,2}$
Ryusuke Jinno,$^{3}$ Kohei Kamada,$^{2,3}$ Seong Chan Park,$^{4}$
} 
\\[3mm]
{\large 
Alexei A.~Starobinsky,$^{2,5}$ and Jun'ichi Yokoyama$^{1,2,6}$
}

\vskip 6mm

$^{1}${\em Department of Physics, Graduate School of Science, }
\\[1mm]
{\em The University of Tokyo, Tokyo 113-0033, Japan}
\\[2mm]
$^{2}${\em Research Center for the Early Universe (RESCEU), Graduate School of Science,}
\\[1mm]
{\em The University of Tokyo, Tokyo 113-0033, Japan}
\\[2mm]
$^{3}${\em Center for Theoretical Physics of the Universe, Institute for Basic Science (IBS),}
\\[1mm]
{\em Daejeon 34126, Korea}
\\[2mm]
$^{4}${\em Department of Physics and IPAP, Yonsei University, Seoul 120-749, Korea}
\\[2mm]
$^{5}${\em L.~D.~Landau Institute for Theoretical Physics, Moscow 119334, Russia}
\\[2mm]
$^{6}${\em Kavli Institute for the Physics and Mathematics of the Universe (Kavli IPMU),}
\\[1mm]
{\em WPI, UTIAS, The University of Tokyo, Kashiwa, Chiba 277-8568, Japan}

\end{center}
\vskip 4mm

\begin{abstract}
It has been argued that the mixed Higgs-$R^2$ model acts as the UV extension of the Higgs inflation, 
pushing up its cut-off scale in the vacuum close up to the Planck scale. 
In this letter, we study the inflaton oscillation stage after inflation, focusing on the effective mass of the 
phase direction of the Higgs field, which can cause a violent preheating process. 
We find that the ``spikes'' in the effective mass of the phase direction observed in the Higgs inflation 
still appear in the mixed Higgs-$R^2$ model. 
While  the spikes appear above the cut-off scale in the Higgs-only case, 
they appear below the cut-off scale when the model is extended with $R^2$ term
though reheating cannot be completed in the violent particle production regime since the spikes get milder. 
\end{abstract}

\end{titlepage}

\tableofcontents
\thispagestyle{empty}
\newpage
\renewcommand{\thefootnote}{$\diamondsuit$\arabic{footnote}}
\setcounter{page}{1}
\setcounter{footnote}{0}

\section{Introduction}
\label{sec:Intro}
\setcounter{equation}{0}

Among many possible candidates of the scalar field that drove inflation in the 
early Universe (see e.g.\ \cite{Sato:2015dga} for review), the Higgs field in the Standard Model (SM)  
${\cal H}$ occupies a unique position as it is the sole (possibly) fundamental scalar field that has actually 
been detected by experiments~\cite{Aad:2012tfa,Chatrchyan:2012xdj}. 
Among many possibilities of Higgs inflation as summarized in \cite{Kamada:2012se}, the original
Higgs inflation model
with a nonminimal coupling to 
gravity, $\xi |{\cal H}|^2 R$ with $\xi = {\cal O}(10^4)$~\cite{CervantesCota:1995tz,Bezrukov:2007ep,Barvinsky:2008ia}, 
is an intriguing model 
because it is  embedded in a simple scale invariant extension of the SM under general relativity
and consistent with the cosmological observations~\cite{Akrami:2018odb}. 
 
However, the quantum mechanical validity of the model had been questioned 
since the Hubble scale during inflation, $H \sim \lambda^{1/4} M_{\rm pl} /\xi^{1/2}$, where $\lambda\simeq 0.01$ and $\mpl$ are the Higgs quartic coupling 
at the inflationary scale and the reduced Planck scale, is 
much higher than the tree-level cut-off scale of the theory in the vacuum, $\Lambda \sim M_{\rm pl}/\xi$~\cite{Burgess:2009ea,Barbon:2009ya,Burgess:2010zq,Hertzberg:2010dc}. 
This issue was resolved as it was discovered that the perturbative cut-off scale of  
interactions of fluctuations around the inflationary background is 
larger than the one in the vacuum~\cite{Barvinsky:2009ii,Bezrukov:2010jz}.
As a result, the background evolution as well as  cosmological perturbations 
generated during inflation is well under control in a quantum mechanical sense\footnote{Quantum
stability during inflation have been shown in \cite{Kamada:2015sca} for
Higgs G-inflation \cite{Kamada:2010qe} and in \cite{Kunimitsu:2015faa} for generalized Higgs inflation.}. 
Nevertheless, due to the nonrenormalizable nature of 
the gravitational coupling in General Relativity (GR), the potential of the Higgs field at the inflationary scale 
cannot be determined by low-energy observables without ambiguity\footnote{Even if the electroweak vacuum is metastable, 
it turns out that inflation can take place with the help of these ambiguities~\cite{Bezrukov:2014ipa}.}~\cite{Bezrukov:2009db,Bezrukov:2014ipa}, 
which requires some UV extension for the complete understanding of the model. 

In fact,  a sensible UV extension is called for even more seriously  to describe the reheating process for the following reason.
Previously, they had been studied in 
e.g. Refs.~\cite{Bezrukov:2008ut,GarciaBellido:2008ab,Repond:2016sol}
following the traditional procedure for inflationary models in GR
\cite{Dolgov:1982th,Abbott:1982hn,Kofman:1994rk,Shtanov:1994ce}, 
and it had been recognized that the depletion of the inflaton quanta is dominated by the 
nonperturbative production of the transverse mode of weak gauge bosons. 
However, it has been recently shown that the effective mass, $m_\theta$, of the 
phase direction of the inflaton or the Nambu-Goldstone (NG) mode, which would constitute  the
longitudinal mode of gauge bosons, exhibits a peculiar behavior 
~\cite{DeCross:2015uza,JinnoThesis,Ema:2016dny}.
It has been shown that violent particle production~\cite{Ema:2016dny,Sfakianakis:2018lzf} 
takes place due to the spiky feature of the effective mass with a large amplitude
$m_{\theta_c}^\mathrm{sp}\simeq \sqrt{\lambda}\mpl $ in a short time scale $\Delta t \simeq (\sqrt{\lambda} M_{\rm pl})^{-1}$.
This is caused by the nonminimal coupling of the Higgs field to gravity in the Jordan frame or by the nontrivial structure of the kinetic term in the Einstein frame, 
but thanks to the conformal duality between inflationary models in the two frames, the physics is identical in either frame. It has been argued
that the mass spikes can cause violent preheating 
in which the NG modes or the longitudinal modes with the momentum $k \simeq \sqrt{\lambda} M_{\rm pl}$ are excited, 
so that most of the energy density of the inflaton can be transferred to the NG modes 
just in one oscillation of the inflaton, which can be the main channel for the depletion of the inflaton quanta~\cite{Ema:2016dny}. 
However, the energy scale of these excitations is far beyond the cutoff scale of the theory during reheating, 
which is much smaller than the one during inflation and already comparable to that in the vacuum, 
$\Lambda \simeq M_{\rm pl}/\xi$, 
for the non-critical Higgs inflation with $\xi\simeq 10^4$ and $\lambda \simeq 0.01$. 
Therefore, such excitations are not quantum mechanically under control and it is not quite clear if the production of the longitudinal modes 
really happens. In order to understand the issue more clearly, 
we need to investigate the behavior of the NG mode at the reheating epoch 
with an appropriate UV extension of the model.

There has been several proposals to push the cutoff scale of the Higgs inflation model
up to the Planck scale~\cite{Giudice:2010ka,Barbon:2015fla,Lee:2018esk}. 
We focus on the mixed Higgs-$R^2$ model~\cite{Wang:2017fuy,Ema:2017rqn,He:2018gyf,Gundhi:2018wyz}, 
where the inflation is driven by the Higgs field and the scalaron from the $R^2$ term~\cite{Starobinsky:1980te},  
which can be also considered as a UV-extension of the Higgs inflation~\cite{Ema:2017rqn,Gorbunov:2018llf}. 
Indeed, the mixed Higgs-$R^2$ model is remarkable, since the new scalar degree of freedom, the scalaron,  naturally arises in it 
as a result of the minimal purely geometric extension of GR without ghosts which makes gravity classically scale invariant for large 
values of the Ricci scalar $R$. 

In this letter, we investigate the behavior of the imaginary part of the Higgs field 
at the reheating epoch in the mixed Higgs-$R^2$ inflation model. 
We show numerically and analytically that the spike gets weakened when the $R^2$ term is included, 
and that the corresponding energy scale becomes lower than the cut-off scale.
As a result, the framework is now under quantum mechanical control as it is desired. 
We also find that the violent preheating is not sufficient to complete the reheating in this case. However, this does not present any
problem for the viability of  Higgs-$R^2$ model since the complete decay of the scalaron and subsequent thermalization can be well 
achieved in the slow perturbative (weak narrow parametric resonance) regime, as it occurs in the pure $R+R^2$ 
model~\cite{Starobinsky:1980te,Starobinsky:1981vz,Vilenkin:1985md,DeFelice:2010aj,Bezrukov:2011gp} due to the effect of gravitational particle creation 
by fast and large oscillations of $R$ in the dust-like post-inflationary epoch.\footnote{Note that the quantitative analysis of creation of matter 
after inflation and the resulting transition of the Universe to the radiation dominated stage in the $R+R^2$ model was performed even earlier 
than that for inflationary models based on GR.}
Although we take a global U(1) for the Higgs field in order to clarify the role of the NG modes at reheating, 
we expect that the conclusion remains unchanged for the fully gauged SU(2)$_L \times$ U(1)$_Y$ case.
Finally, throughout this letter, we assume that the Higgs potential is completely stable 
and $\lambda \simeq 0.01$ at the inflationary scales. For other realizations of Higgs inflation, the preheating dynamics can be 
completely different. In the critical Higgs inflation, since $\lambda \ll 1$ and $\xi\sim 10^5\sqrt{\lambda}\sim O(1)$, 
we expect that violent spikes do not appear and 
there exists only a single cut-off scale at the Planck scale~\cite{Hamada:2014iga,Bezrukov:2014bra,Hamada:2014wna}.
Smaller values of $\xi$ ($\ll 10^4$) are also possible in the hilltop case~\cite{Enckell:2018kkc}.
In the hillclimbing Higgs inflation~\cite{Jinno:2017jxc,Jinno:2017lun}, strong spikes can emerge,
though no preheating analysis has been performed yet.

\section{Mixed Higgs-$R^2 $ Model}
\label{sec:L-EOM}
\setcounter{equation}{0}

\subsection{Action}

Let us briefly review the structure and inflationary dynamics of the mixed Higgs-$R^2$ 
model~\cite{Wang:2017fuy,Ema:2017rqn,He:2018gyf,Gundhi:2018wyz}.
We start from the action in the Jordan frame with a complex scalar ${\mathcal H}$ 
nonminimally coupled to the Ricci scalar and the $R^2$ term
\begin{align}
S_{\mathrm J}
= \int\!\! d^4 x \sqrt{-g_{\mathrm J}} {\cal L}_\mathrm{J}=
\int\!\! d^4 x \sqrt{-g_{\mathrm J}} 
\left[ \left(\frac{M_{\mathrm{pl}}^2}{2} + \xi |{\mathcal H}|^2 \right)R_{\mathrm J} + \frac{M_{\mathrm{pl}}^2}{12 M^2}R_{\mathrm J}^2 
- g_{\mathrm J}^{\mu\nu} \partial_\mu  {\mathcal H} \partial_\nu {\mathcal H}^\dagger - \lambda |{\mathcal H}|^4 \right],  \label{jordanaction}
\end{align}
where $M$ is a parameter with a mass dimension one, which will be identified as the scalaron mass for low $R$ (in particular, in flat space-time).
The subscript J represents that the variables are the ones in the Jordan frame, 
and we will use the subscript E for the Einstein frame. 
We would like to regard ${\mathcal H}$ as the SM Higgs, but in order to make the argument simple and explicit, 
we first take a toy model with a global U(1) symmetric scalar, without introducing a gauge field. Still, we expect that the results remain unchanged for the SM Higgs field
charged under gauged SU(2)$_L \times$ U(1)$_Y$. 
Here we take the sign convention where the metric is taken to be $g_{\mu\nu} = (-,+,+,+)$ at the flat limit and the nonminimal coupling is $\xi=-1/6$ in the case of the conformally coupled scalar. In order for inflation driven by the ${\mathcal H}$ field to occur, we consider the case $\xi>0$.

Defining the scalaron field $\varphi$ as \cite{Maeda:1988ab,Maeda:1987xf} 
\begin{align}
   \label{def:psi}\sqrt{\frac{2}{3}}\frac{\varphi}{\mpl}\equiv \ln \lmk\frac{2}{\mpl^2}\left|\frac{\partial {\cal L}_\mathrm{J}}{\partial {R_\mathrm{J}}}\right|\rmk,
\end{align}
and performing a conformal transformation
\begin{align}
   g_{\mathrm{E}\mu \nu} (x)=e^{\sqrt{\frac{2}{3}}\frac{\varphi(x)}{\mpl}} {g_\mathrm{J}}_{\mu \nu}(x)
   \equiv e^{\alpha\varphi(x)}{g_\mathrm{J}}_{\mu \nu}(x),
\end{align}
we can transform the original action \eqref{jordanaction} into the one in the
Einstein frame with two scalar fields, $\varphi$ and ${\mathcal H}$ and express the new action in terms of the new scalar fields as
\begin{align}
\label{einsteinaction}
S_{\mathrm{E}} 
&=
\int d^4 x \sqrt{-g_{\mathrm E}} 
\left[\frac{M_{\mathrm{pl}}^2}{2} R_{\mathrm{E}} -\frac{1}{2} g_{\mathrm E}^{\mu\nu}\partial_\mu \varphi \partial_\nu \varphi 
- e^{-\alpha \varphi} g^{\mu\nu}_{\mathrm E} \partial_\mu {\mathcal H} \partial_\nu {\mathcal H}^\dagger  -U(\varphi, {\mathcal H}) \right],
\\ 
\label{potential}
U(\varphi, {\mathcal H}) 
&= 
\lambda e^{-2 \alpha \varphi}| {\mathcal H}|^4  + \frac{3}{4} M_{\mathrm{pl}}^2 M^2 
\left[1 - \left(1 + \frac{2\xi}{M_{\mathrm{pl}}^2} |{\mathcal H}|^2\right)e^{-\alpha \varphi} \right]^2. 
\end{align}

Hereafter we study the system in the Einstein frame, but the physical results are the same when we study it in the Jordan frame (though actual values of space-time curvature and particle energies are different).
The potential terms contain higher dimensional operators as well as induced quartic couplings 
which prevents us from performing quantum analysis up to arbitrary high energy scales. 
The perturbativity of the system around the origin $\varphi \simeq {\mathcal H} \simeq 0$ with respect to the Higgs field is determined 
by the effective coupling for $|{\mathcal H}|^4$, which yields an upper bound for the scalaron mass as~\cite{Ema:2017rqn,Gorbunov:2018llf}
\begin{align}
M
&\lesssim 
\sqrt{\frac{4 \pi}{3}} \frac{M_{\mathrm{pl}}}{\xi}.  \label{pertcond}
\end{align}
Once this condition is satisfied, the perturbative cut-off scales of other higher order interactions becomes larger than the reduced Planck 
mass, so that the cut-off scale of the system is identified as $\Lambda \simeq M_{\mathrm{pl}}$. Note that the condition (\ref{pertcond}) 
produces no significant new bound in the small coupling case $\xi \lesssim 1$ including $\xi = 0$.

\subsection{Inflationary dynamics}

The classical dynamics of the system is determined by the scalaron $\varphi$ and the radial direction of the `Higgs' field, 
$h$, defined by $\cH=h e^{i\theta}/\sqrt{2}$ where $\theta$ represents the Nambu-Goldstone mode which constitutes 
a longitudinal mode of the gauge fields in a more realistic theory.
For each value of $\varphi>0$,  the potential along $h$ direction is minimized at  \cite{Ema:2017rqn}
\begin{align}
h^2
&=
\frac{e^{\alpha \varphi} - 1}{\displaystyle \frac{\xi}{\mpl^2} + \frac{\lambda}{3\xi M^2}}
\label{valley}
\end{align}
that corresponds to $h^2=\xi R_{\text{J}}/\lambda$ in the Jordan frame.
Because of the non-flat metric in the field space as observed in the kinetic terms in (\ref{einsteinaction}), the 
location of the valley of the potential is slightly shifted from those given by (\ref{valley}) \cite{Wang:2017fuy}.
Furthermore, the actual dynamics does not trace either the local minimum of the potential along $h$ direction
nor the valley as shown in the most comprehensive analysis presented in \cite{He:2018gyf}.

Fortunately, however, as far as observable quantities such as the amplitude and the spectral index
of the curvature perturbation are concerned, we may use the approximate relation (\ref{valley}) to study
the dynamics during inflation in terms of $ \varphi $~\cite{He:2018gyf} as long as $ \lambda $ is not too small~\cite{Wang:2017fuy,Gundhi:2018wyz}. Inserting (\ref{valley}) in (\ref{einsteinaction}) we 
find that both the kinetic term and the potential take an equivalent form as the pure Higgs inflation model
in the Einstein frame with modified effective coupling constants
\begin{align}
\tilde{\lambda}
&\equiv 
\lambda \left(1 + \frac{\lambda M_\mathrm{pl}^2}{3 \xi^2 M^2}\right), 
\quad
\tilde{\xi}
\equiv
\xi \left(1+\frac{\lambda M_\mathrm{pl}^2}{3 \xi^2 M^2}\right), 
\end{align}
with the potential energy density in the plateau region given by
\begin{align}
U_\mathrm{inf} 
&=
\frac{\tilde{\lambda} M_\mathrm{pl}^4}{4 \tilde{\xi}^2} 
= 
\frac{\lambda M_\mathrm{pl}^4}{4\xi^2 \left(1+\dfrac{\lambda M_\mathrm{pl}^2}{3 \xi^2 M^2}\right)}. 
\label{infpot}
\end{align}

On the other hand, we can also obtain an effective $R^2$ theory starting from the action in the Jordan frame (\ref{jordanaction}) 
by neglecting the Higgs kinetic term which is a good approximation when $\xi$ is much larger than unity \cite{He:2018gyf}.
In this case the field equation of the Higgs yields a constraint $h^2=\xi R_\mathrm{J}/\lambda$ so that
the action reduces to
\begin{align}
S_{\text{J}} 
&=
\int d^4 x \sqrt{-{g_\mathrm{J}}} \left[\frac{\mpl^2}{2} {R_\mathrm{J}}+\left(\frac{\mpl^2}{12 M^2}+\frac{\xi^2}{4\lambda}\right) R_\mathrm{J}^2 \right] 
\nonumber \\
&=
\int d^4 x \sqrt{-{g_\mathrm{J}}} \left[\frac{\mpl^2}{2} {R_\mathrm{J}}+\frac{\mpl^2}{12 {\tilde M}^2}R_\mathrm{J}^2 \right],
\label{r2action}
\end{align}
where 
\begin{align}
\tilde{M}^2
&\equiv 
\frac{M^2}{\displaystyle 1 + \frac{3\xi^2M^2}{\lambda \mpl^2}}
\label{meff:scalaron}
\end{align}
is the effective mass squared of the scalaron.
If we transform the effective action (\ref{r2action}) to the Einstein frame we obtain the 
well-known form of the scalaron potential with the potential height $U_\mathrm{inf}=3\mpl^2 \tilde{M}^2/4$
in the plateau region, which is to be identified with (\ref{infpot}).

\begin{figure}[h]
\centering
\includegraphics[width=.45 \textwidth]{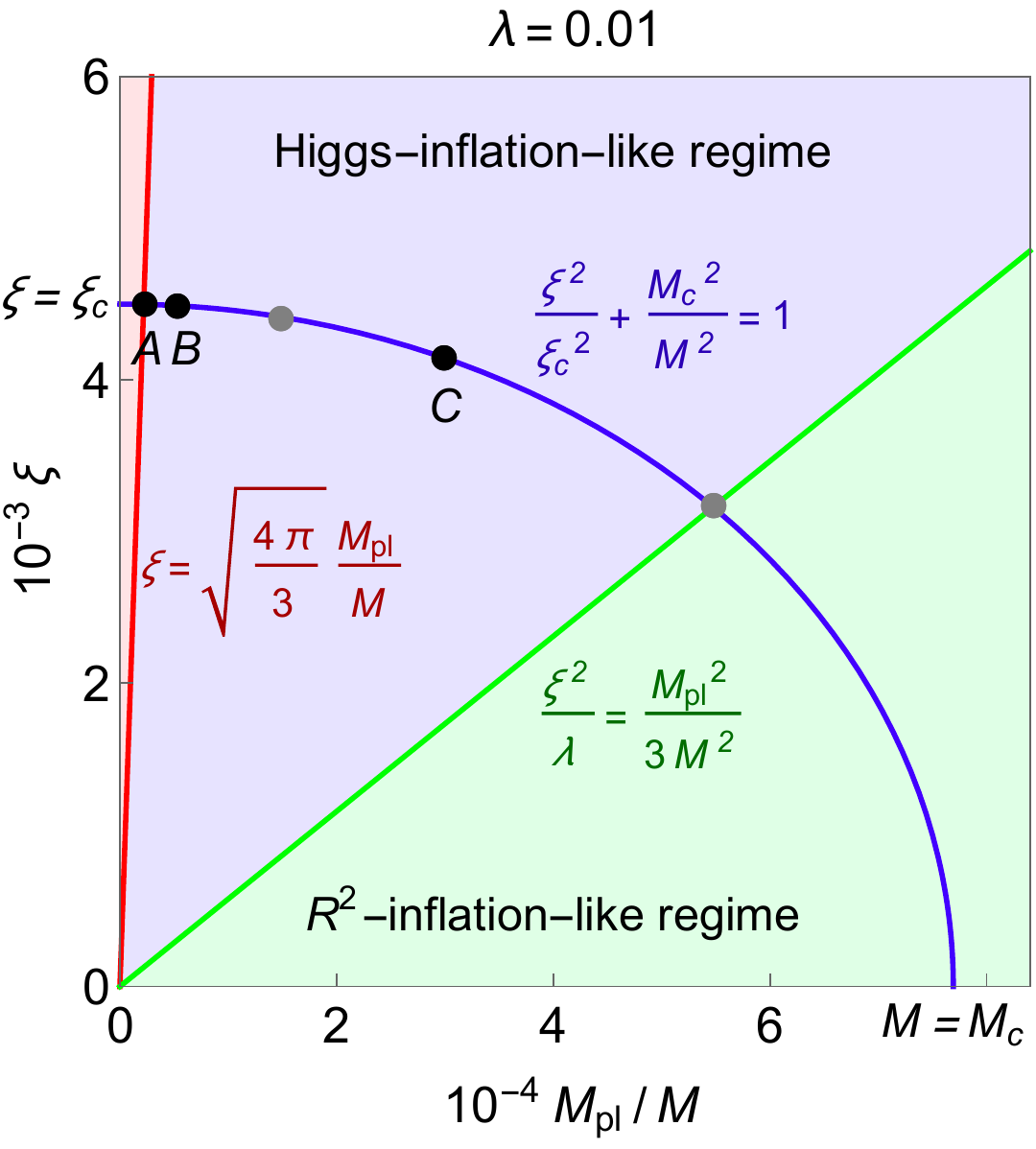}
\caption{
Parameter space for different regimes in the mixed Higgs-$R^2$ model with $\lambda = 0.01$. 
The red region is the strongly-coupled regime where perturbative analysis is questionable~\eqref{pertcond}. 
The blue and green regions are the Higgs-like and $R^2$-like regimes, respectively.
The blue line satisfies the condition for the observed curvature power spectrum \eqref{eq:Mxiconstraint}.
The parameter points A, B, and C along the blue line represent our benchmark points: see Sec.~\ref{subsec:Background}.
The black and gray points are the benchmark points for Figs.~\ref{fig:FieldEvolution}--\ref{fig:MassEvolution}.
}
\label{fig:ParameterSpace}
\end{figure}

The quantities $\tilde{\lambda}, \tilde{\xi}$, and $\tilde{M}$ 
are determined by the amplitude of the curvature perturbation
${\mathcal P}_{\mathcal R} \simeq 2.1 \times 10^{-9}$~\cite{Akrami:2018odb} 
at the pivot scale which left the Hubble horizon $N$ $e$-folds before the end of inflation. 
We find \cite{Starobinsky:1983zz,Faulkner:2006ub}
\begin{align}
\frac{\tilde{\xi}^2}{\tilde{\lambda}} 
&= 
\frac{\mpl^2}{3 \tilde{M}^2}
=
\frac{\xi^2}{\lambda} + \frac{M_\mathrm{pl}^2}{3M^2} 
=
\frac{N^2}{72 \pi^2{\mathcal P}_{\mathcal R}}.
\label{eq:Mxiconstrainto}
\end{align}
Defining $\xi_c$ and $M_c$ as
\begin{align}
\xi_c
&\equiv
\sqrt{\frac{\lambda N^2}{72 \pi^2{\mathcal P}_{\mathcal R}}} \simeq 4.4 \times 10^3 \left(\frac{\lambda}{0.01}\right)^{1/2} \left(\frac{N}{54}\right), \\ 
~~~~
M_c
&\equiv
\sqrt{\frac{24\pi^2 {\mathcal P}_{\mathcal R}}{N^2}} M_\mathrm{pl}
\simeq
1.3 \times 10^{-5}\lmk\frac{N}{54}\rmk^{-1}M_\mathrm{pl},
\end{align}
we see that $\tilde{M}$ is constrained to be $\tilde{M} = M_c$.
In the following we fix $N\simeq 54$ for definiteness. 
We also see that observationally viable mixed Higgs-$R^2$ inflation satisfies
\begin{align}
\frac{\xi^2}{\xi_c^2} + \frac{M_c^2}{M^2}
&= 1.
\label{eq:Mxiconstraint}
\end{align}
From this parametrization, we see the following two limits:
\begin{itemize}
\item
Pure-$R^2$ inflation limit: 
$\xi \ll \xi_c$ and $M \rightarrow M_c$,
\item
Pure-Higgs inflation limit:
$\xi \rightarrow \xi_c$ and $M \rightarrow \infty$.
\end{itemize}
Note that the pure-Higgs inflation limit does not respect the perturbativity condition~\eqref{pertcond}. 
Also, in the whole parameter region, we define the Higgs-like and $R^2$-like regimes as follows:
\begin{itemize}
\item
$R^2$-like regime: 
\begin{align}
\frac{\xi^2}{\lambda}
&< \frac{M_\mathrm{pl}^2}{3M^2},
\end{align}
\item
Higgs-like regime:
\begin{align}
\frac{\xi^2}{\lambda}
&> \frac{M_\mathrm{pl}^2}{3M^2}.
\end{align}
\end{itemize}
Figure~\ref{fig:ParameterSpace} shows the parameter space in the $\xi$-$1/M$ plane with $\lambda=0.01$
for the $R^2$-like regime, Higgs-like regime, and strongly coupled regime~\eqref{pertcond}, which covers only part of the Higgs-like regime for $\lambda < 1$.
Indeed, with the condition \eqref{eq:Mxiconstraint}, the Higgs-like region without the strong coupling issue is given by
$3.1 \times 10^3 \sqrt{\lambda/0.01} \lesssim \xi \lesssim 4.4 \times 10^3 \sqrt{\lambda/0.01}$ 
(or $2 \times 10^3 \sqrt{\lambda/0.01} \lesssim M_\mathrm{pl}/M \lesssim 5.4 \times 10^4$).
Therefore, there exists a parameter space within the Higgs-like region where the system
is quantum mechanically under control up to the Planck scale~\cite{Ema:2017rqn,Gorbunov:2018llf}\footnote{Note for comparison that in the minimally coupled case $\xi=0$, the double inflationary $ h^4$-$R^2$ model was first considered
in~\cite{Kofman:1985aw} without identifying $h$ with the Higgs field, and its scalar perturbation spectrum was derived
in~\cite{Starobinsky:1986fxa}. To obtain the correct value for the slope
of the scalar power spectrum $n_s-1$ and to satisfy the upper limit on the
tensor-to-scalar ratio $r$, viability of such a model requires its last
$\sim 60$ e-folds to be in the $R^2$-like regime that occurs for $M <
\sqrt{\lambda} M_\mathrm{pl}$, or if $h^2$ is always less than $M_\mathrm{pl}^2$ (and
then the field $h$ does not contribute to inflation at all). The same
conclusion remains valid for $0<\xi\ll 1,~\xi h^2/M_\mathrm{pl}^2\ll 1$, too.}.

\section{Inflaton dynamics and preheating after inflation}
\label{sec:Preheating}
\setcounter{equation}{0}

Let us now analyze the inflaton dynamics and its effect on the light direction (i.e. the phase direction)
after inflation.

\subsection{Background evolution}
\label{subsec:Background}

After inflation, both $h$ and $\varphi$ roll down rapidly to the origin and start to oscillate around it
with the equations of motion
\begin{align}
\ddot{\varphi}+3H\dot{\varphi}+\frac{\alpha}{2}e^{-\alpha \varphi}\dot{h}^2+\frac{\partial U}{\partial \varphi}
= 0, \label{eq:eom1}
&\\
\ddot{h}+3H\dot{h}- \alpha \dot{\varphi}\dot{h}+e^{\alpha \varphi}\frac{\partial U}{\partial h} 
= 0, \label{eq:eom2}
&\\
{\rm with}~~~~~~~3M^2_\mathrm{pl}H^2=\frac{1}{2}\dot{\varphi}^2+\frac{1}{2}e^{-\alpha \varphi}\dot{h}^2 + U(\varphi,h). \label{eq:eom3}
&
\end{align}
When the scalaron mass satisfies $M<\sqrt{\lambda} M_\mathrm{pl}$, the effective single-field
description does not apply in the field oscillation regime, and  the trajectory of the scalar fields 
as given by (\ref{eq:eom1}) and  (\ref{eq:eom2})
becomes highly complicated.
Figures~\ref{fig:FieldEvolution} show typical evolution of $h$ (top panels) and  $\varphi$ (bottom panels)
for three benchmark points chosen as follows:
\begin{itemize}
\item[(A)]
$\xi/\xi_c \simeq 0.9996,\ M_c/M \simeq 0.0282
\quad
\leftrightarrow
\quad
\xi \simeq 4439, \ M_\mathrm{pl}/M \simeq 2.17\times 10^3$,
\item[(B)]
$\xi/\xi_c \simeq 0.9975, \ M_c/M \simeq 0.0709
\quad
\leftrightarrow
\quad
\xi=4430, \ M_\mathrm{pl}/M \simeq 5.45\times 10^3$,
\item[(C)]
$\xi/\xi_c \simeq 0.9208, \ M_c/M \simeq 0.39 \ \ \ 
\quad
\leftrightarrow
\quad
\xi \simeq 4089, \ M_\mathrm{pl}/M = 3 \times 10^4$,
\end{itemize}
which satisfy the observational constraint~\eqref{eq:Mxiconstraint}. 
Note that parameter point A lies on the boundary to the strongly-coupled regime.  
Here we take the initial condition just before the end of inflation 
as $\varphi=1.2 M_\mathrm{pl}, {\dot \varphi} = 0$, while $h$ satisfies $\partial U/\partial h =0$ 
and $ {\dot h}=0$ at $t=0$, 
but we have confirmed that our results remain unchanged if we take larger number of $e$-folds before the end of inflation. 
We see that the scalar fields are once trapped in the narrow valley for $\varphi<0$ with the time scale $\Delta t \sim M^{-1}$ 
and the $h$ field oscillates rapidly with the effective mass squared $\sim \xi M^2$ (for $|\varphi| \simeq  M_\mathrm{pl}$) around the stream line at the bottom of the valley.

\subsection{Effective mass for the phase direction}

\begin{figure}
\centering
\includegraphics[width=.32\textwidth]{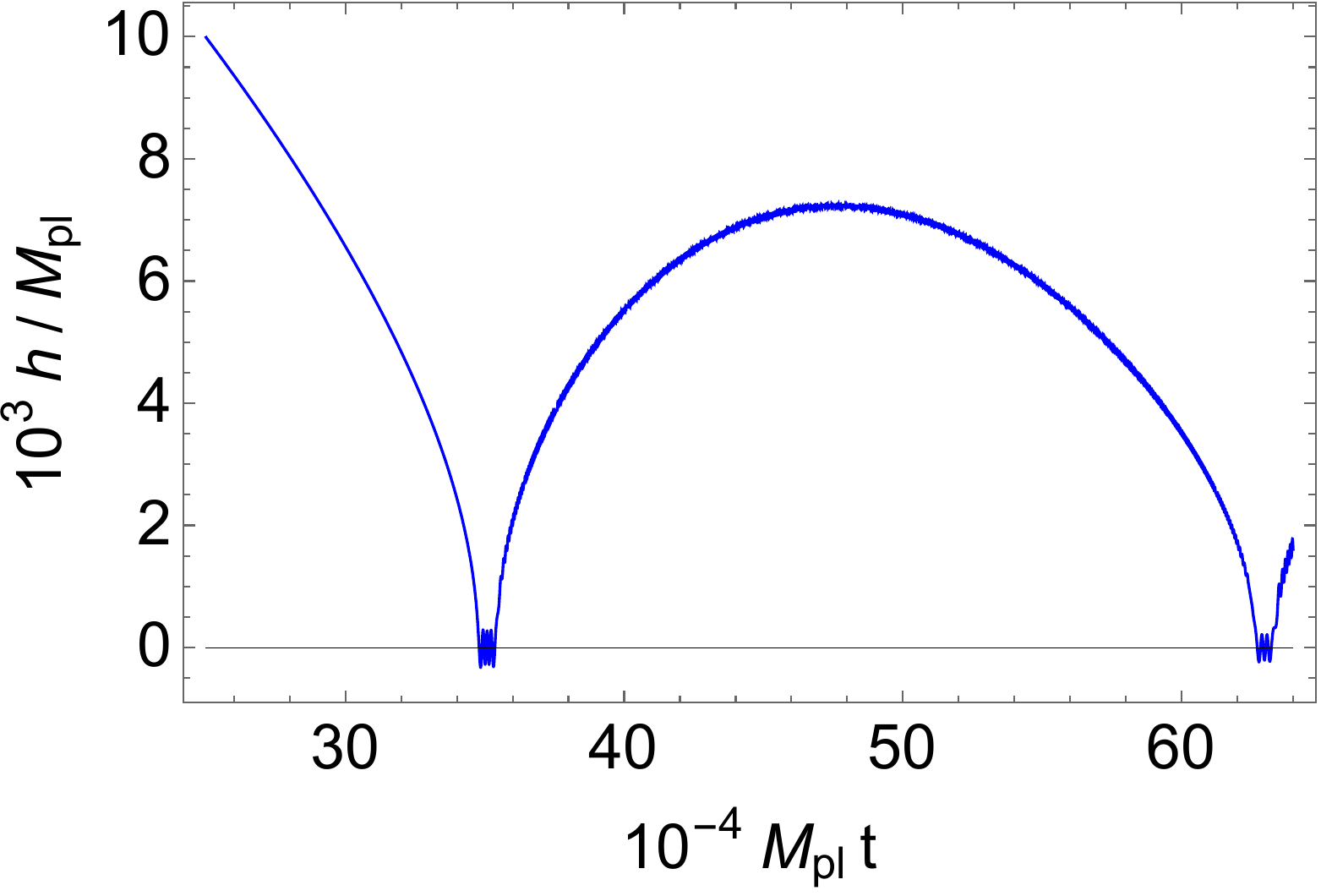}
\includegraphics[width=.32\textwidth]{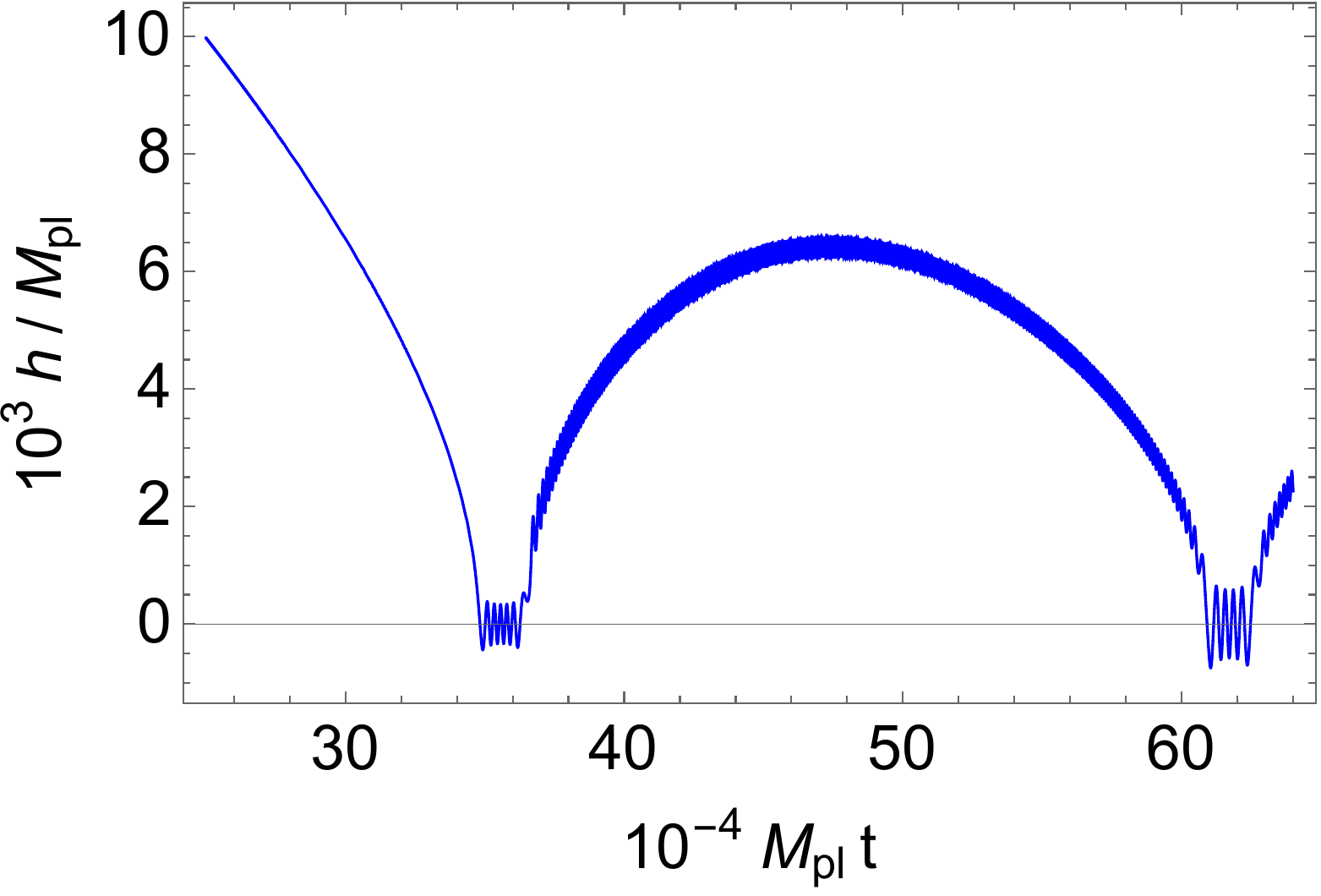}
\includegraphics[width=.32\textwidth]{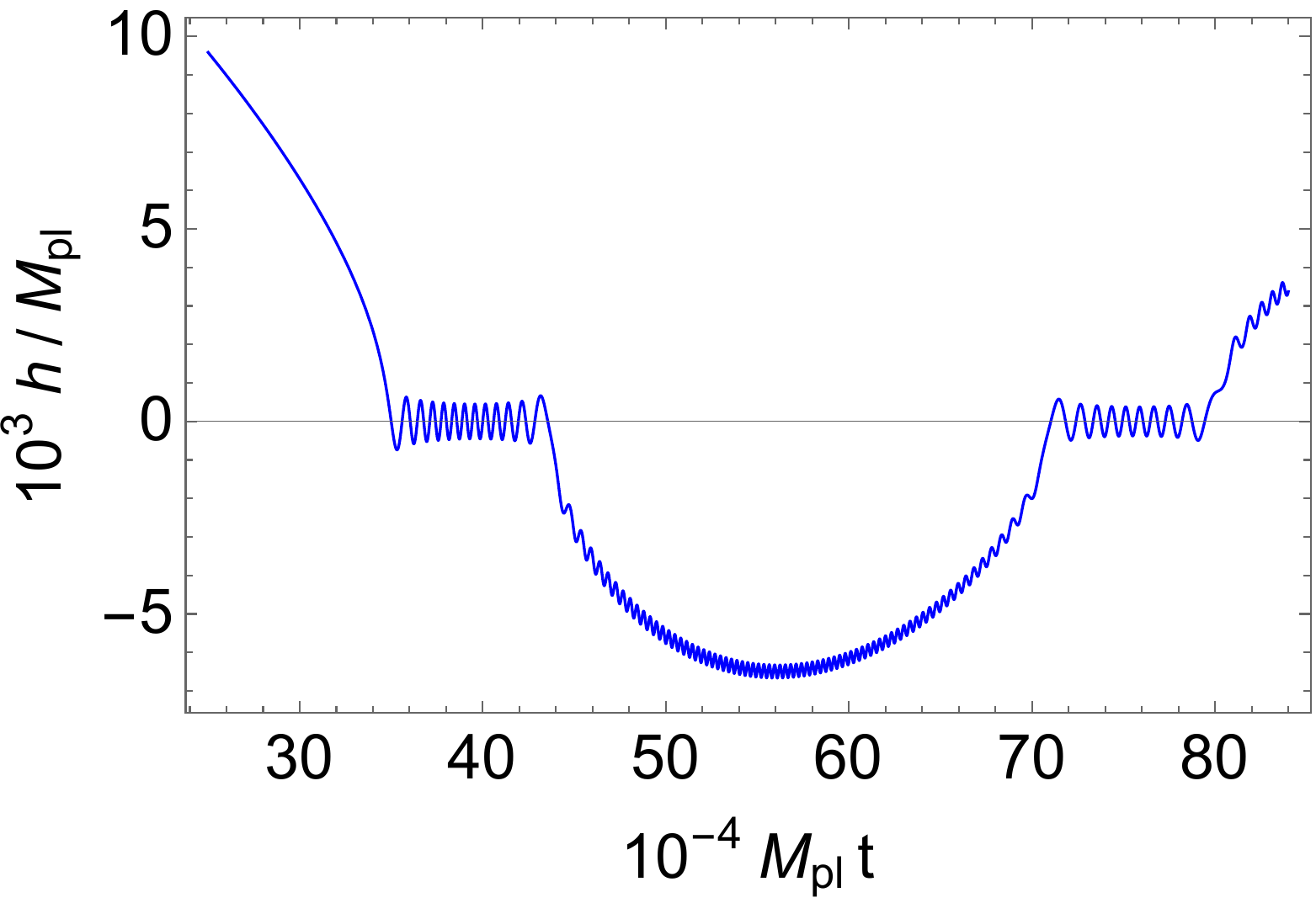}
\includegraphics[width=.32\textwidth]{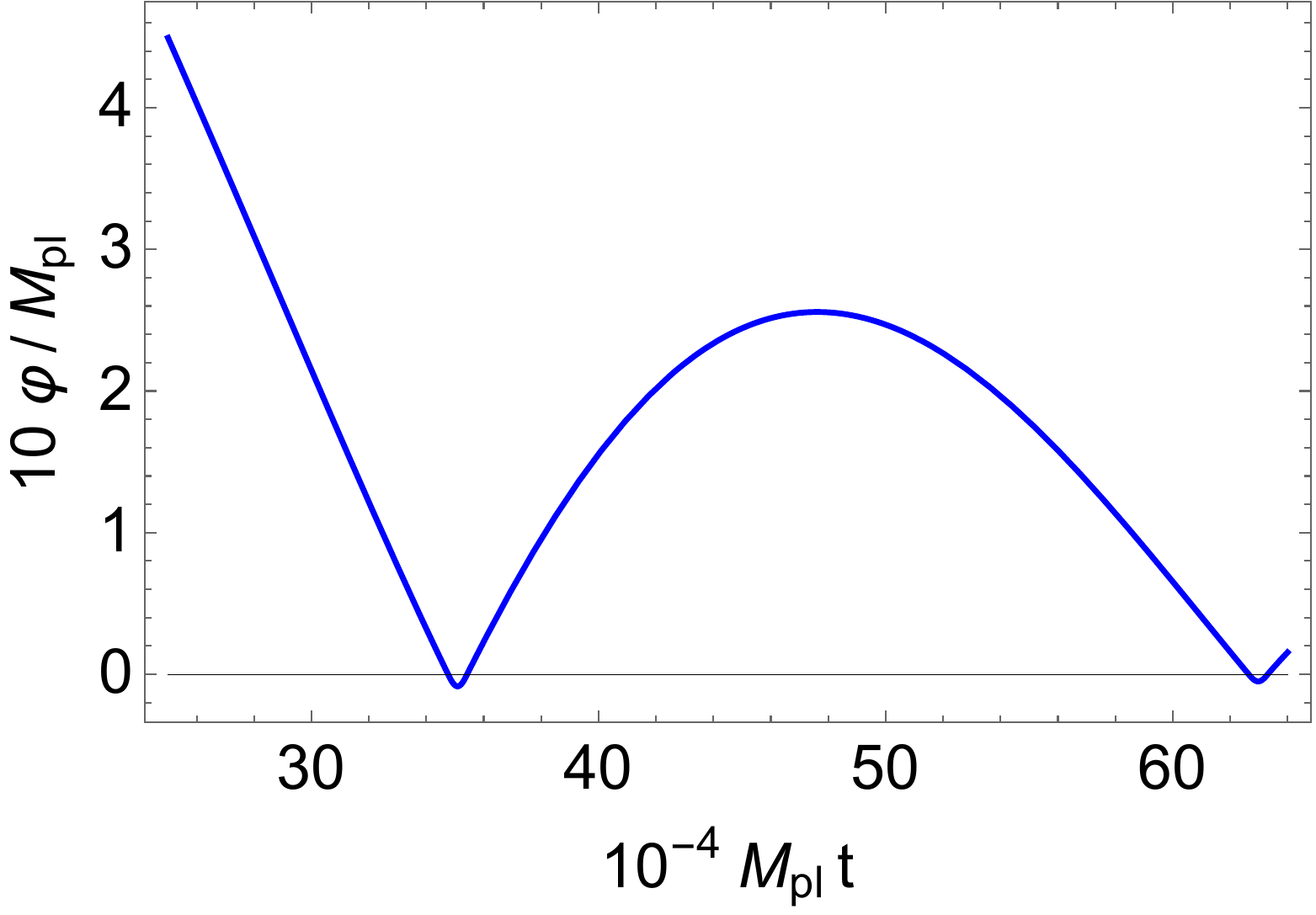}
\includegraphics[width=.32\textwidth]{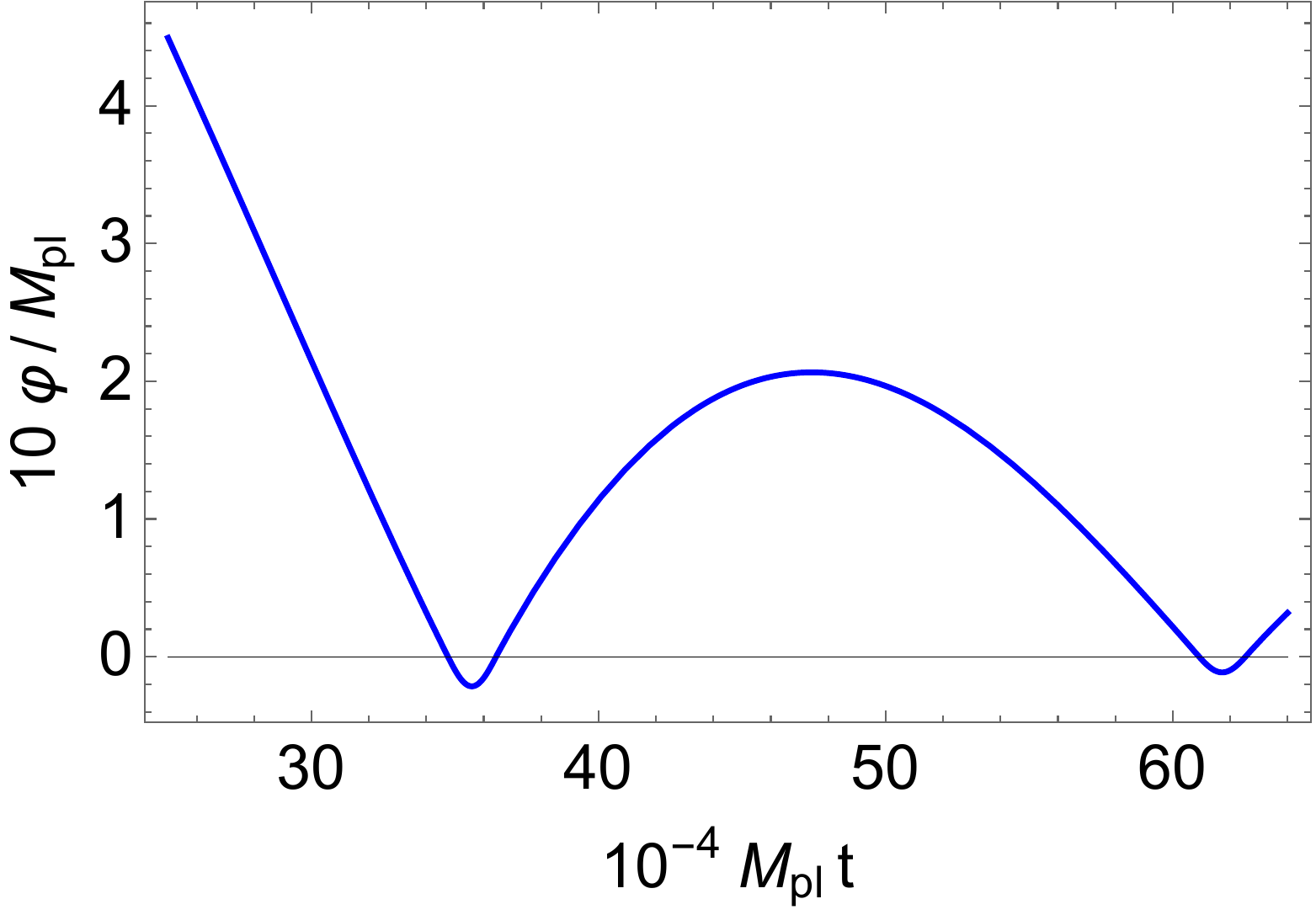}
\includegraphics[width=.32\textwidth]{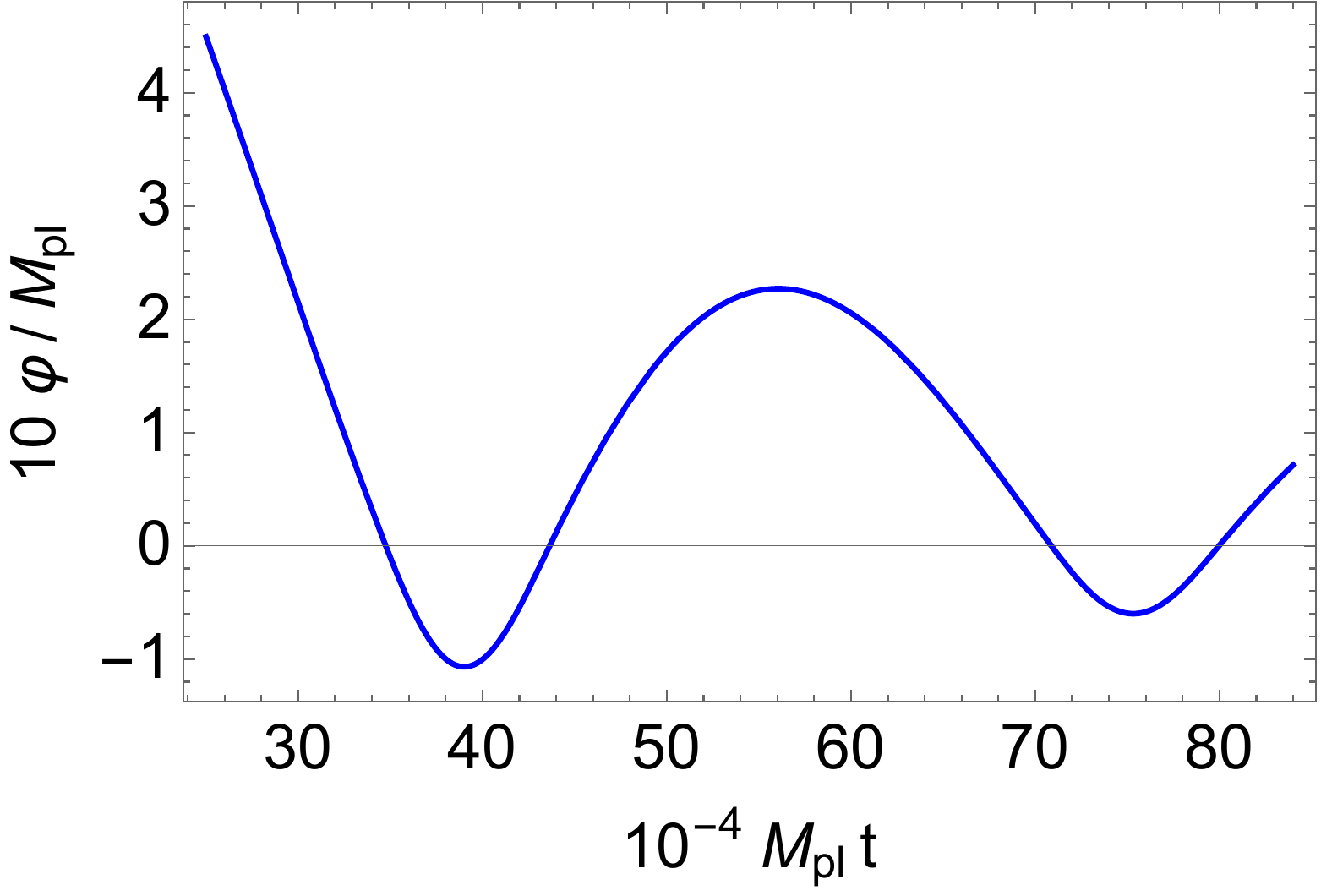}
\caption{
Time evolution of the Higgs field $h$ (top) and scalaron $\varphi$ (bottom) 
for the parameter points (A) (left), (B) (middle), and (C) (right).
We fixed $\lambda=0.01$. See Fig.~\ref{fig:ParameterSpace} for the three parameter points.
}
\label{fig:FieldEvolution}
\vskip 10mm
\includegraphics[width=.32\textwidth]{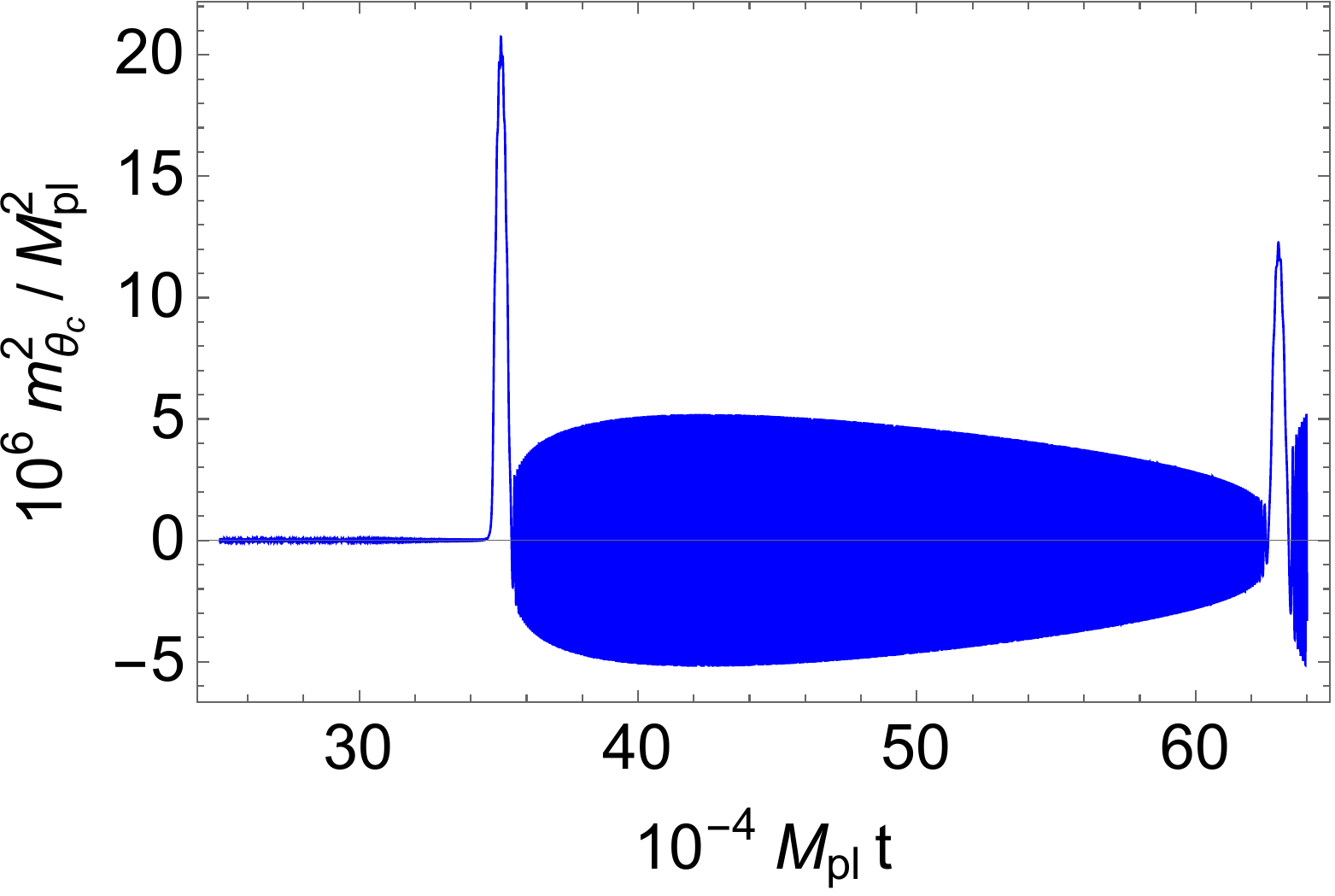}
\includegraphics[width=.32\textwidth]{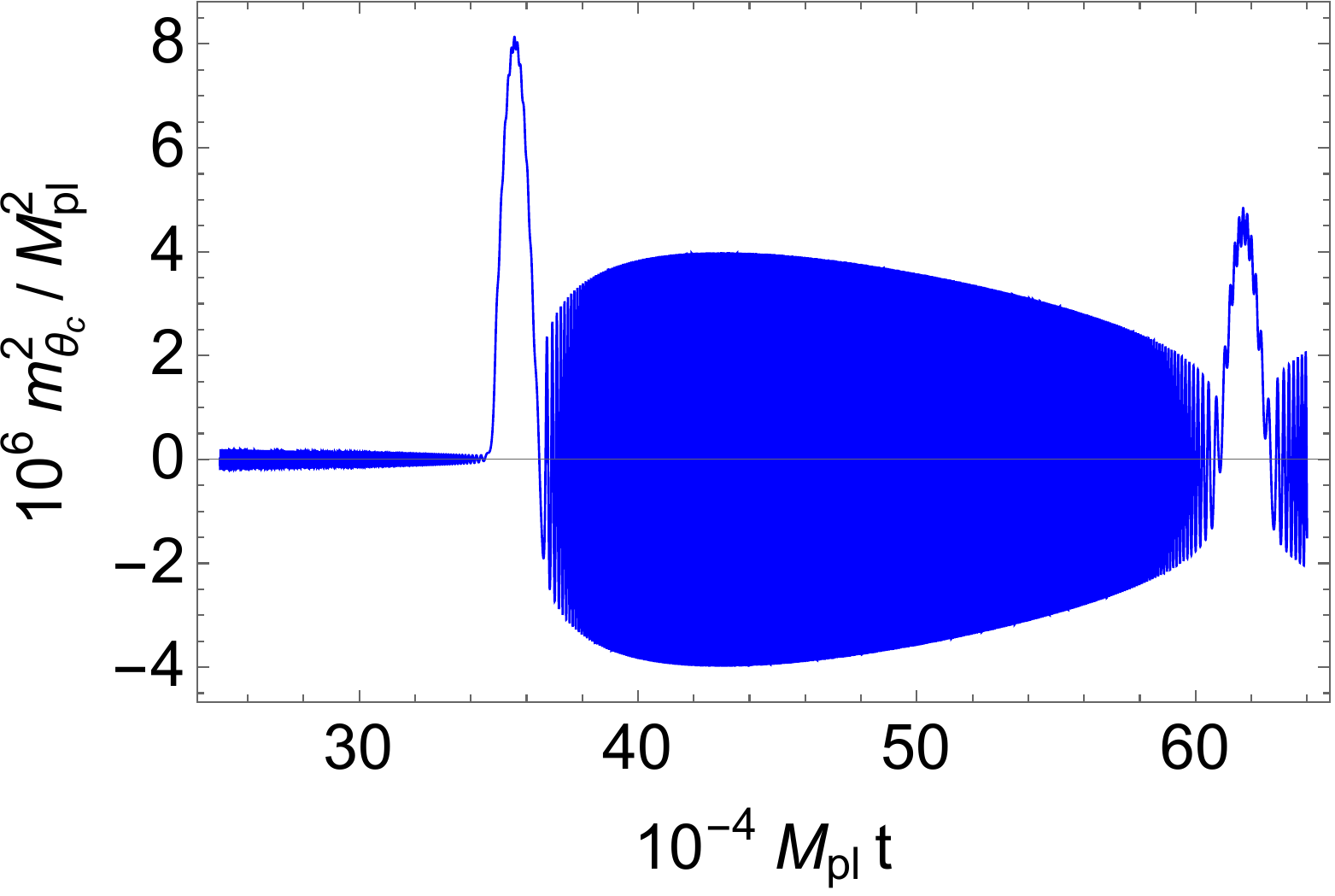}
\includegraphics[width=.32\textwidth]{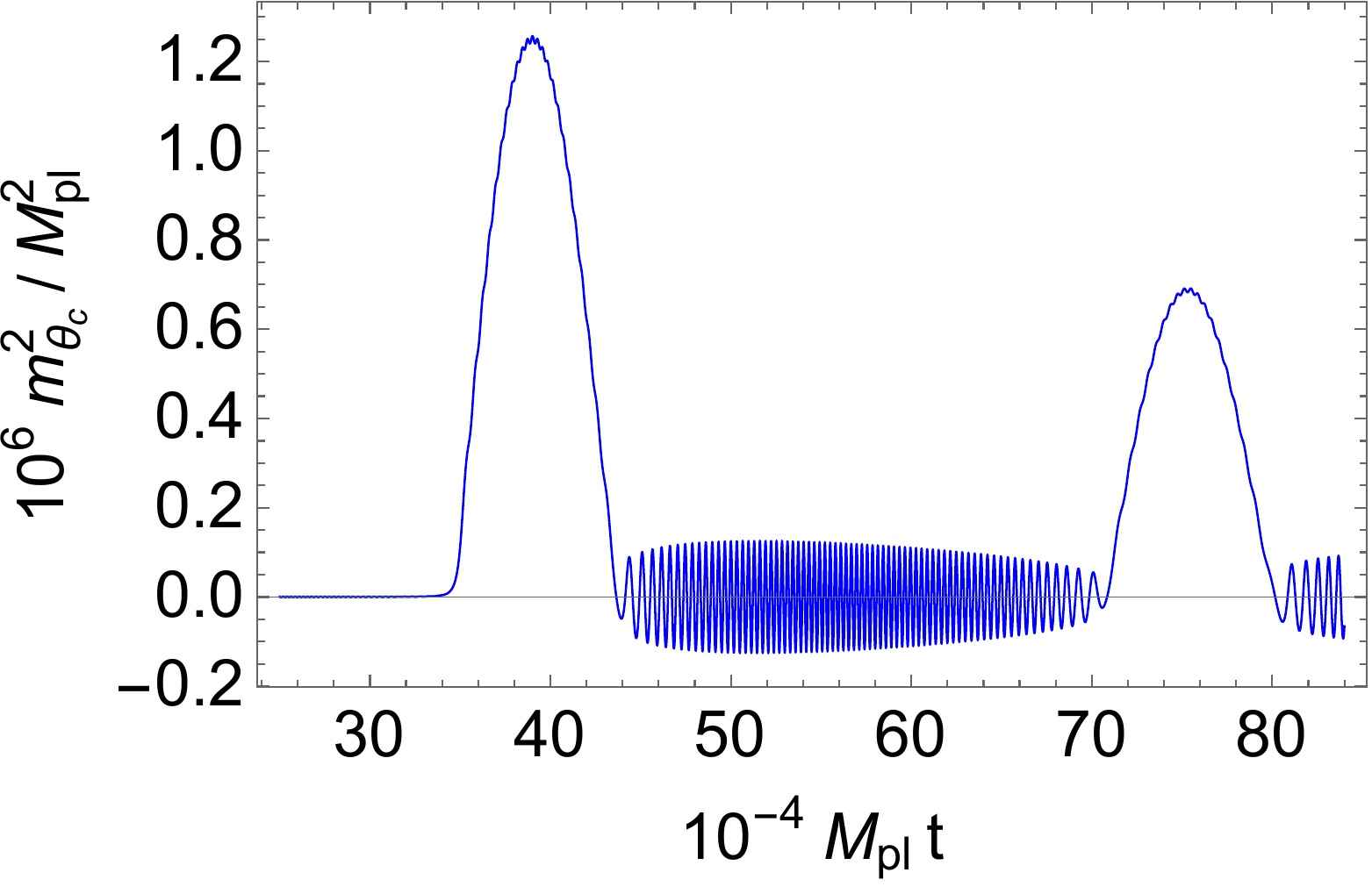}
\includegraphics[width=.32\textwidth]{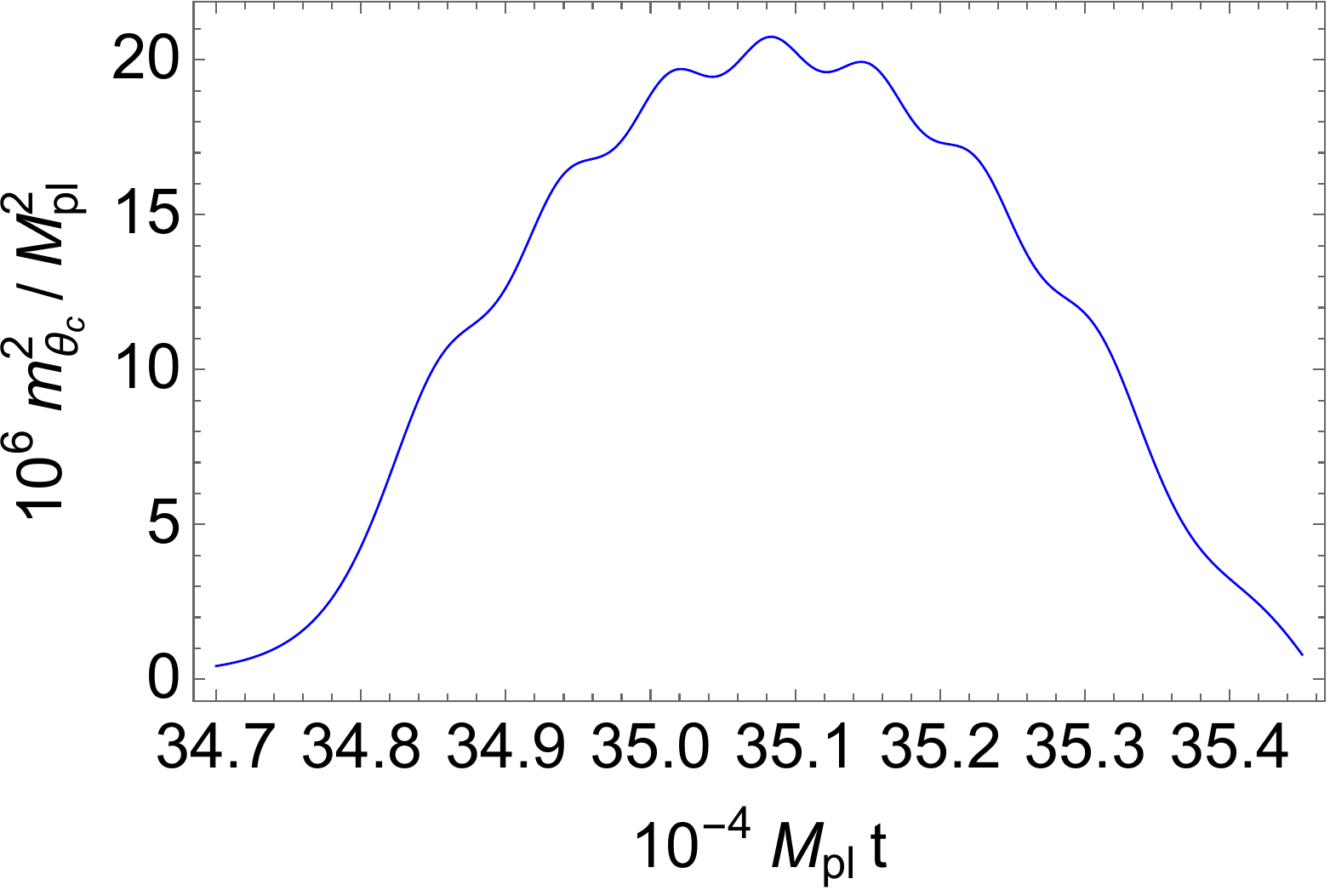}
\includegraphics[width=.32\textwidth]{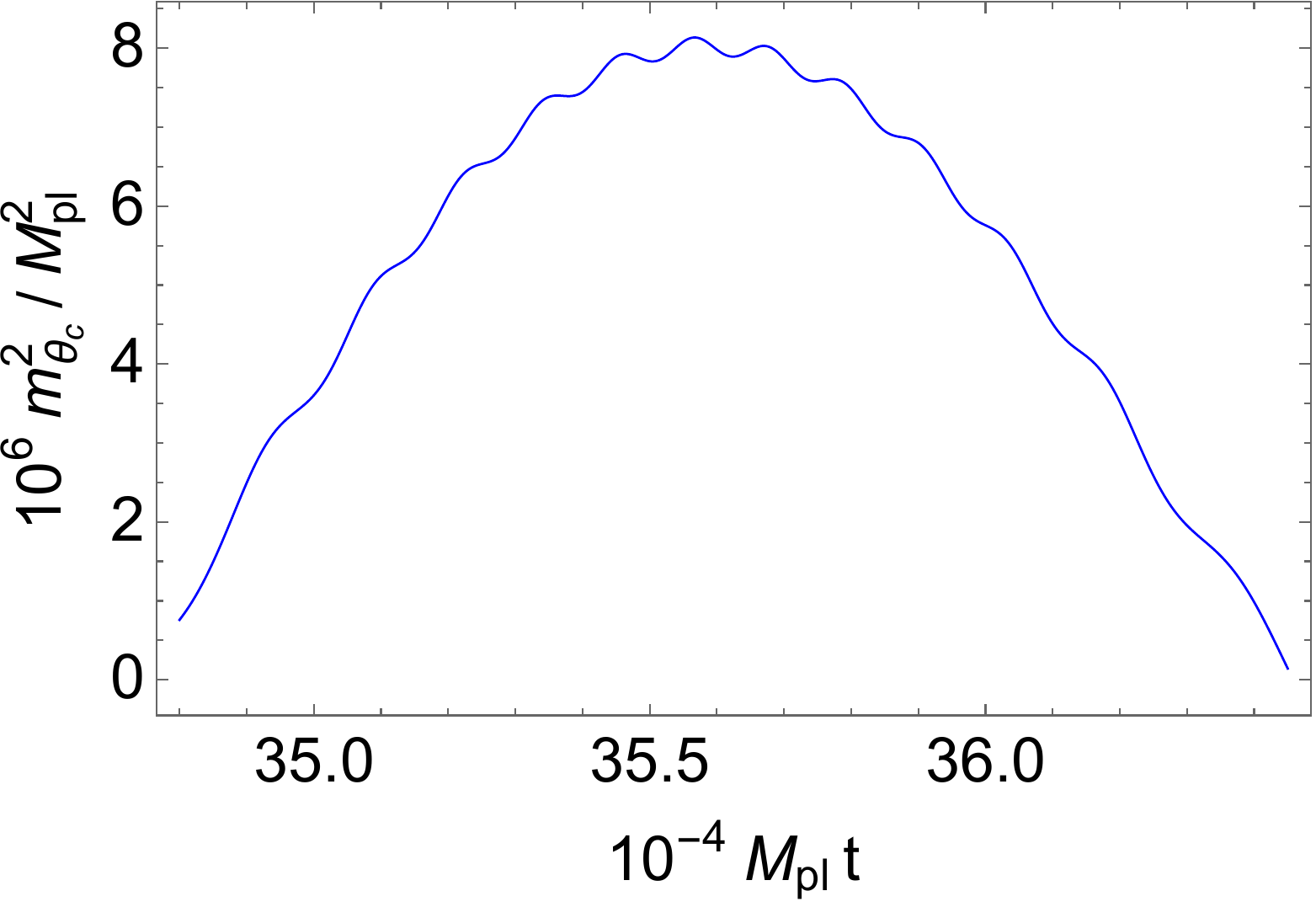}
\includegraphics[width=.32\textwidth]{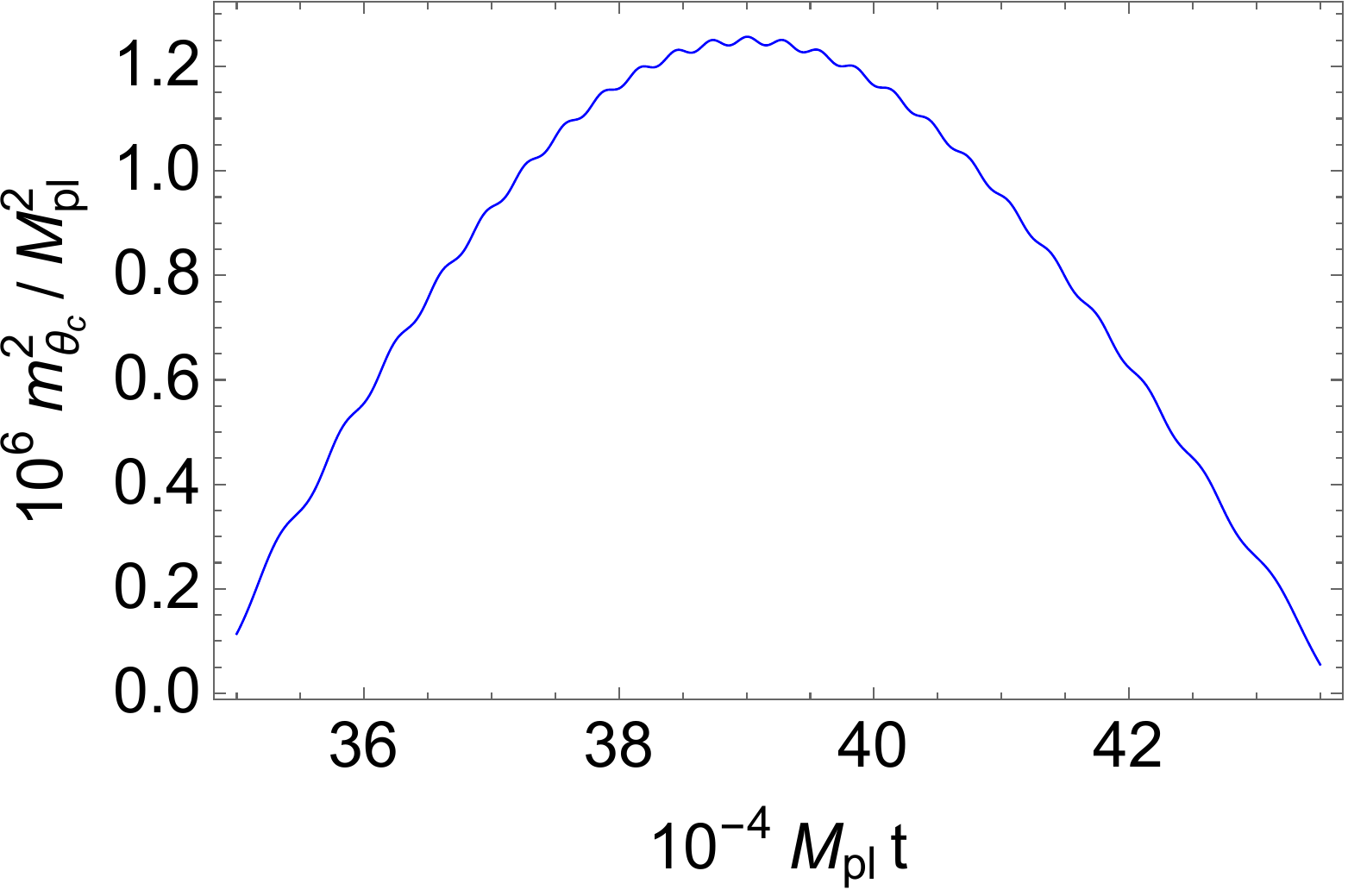}
\caption{
Time evolution of the effective mass squared for the phase direction $m^2_{\theta_c}$
for the parameter points (A) (left), (B) (middle), and (C) (right).
We fixed $\lambda=0.01$.
The top panels show the evolution over the full time range shown in Fig.~\ref{fig:FieldEvolution},
while the bottom panels are magnifications of the top panels around the first peak.
}
\label{fig:MassEvolution}
\end{figure}

\begin{figure}
\centering
\includegraphics[width=.6\textwidth]{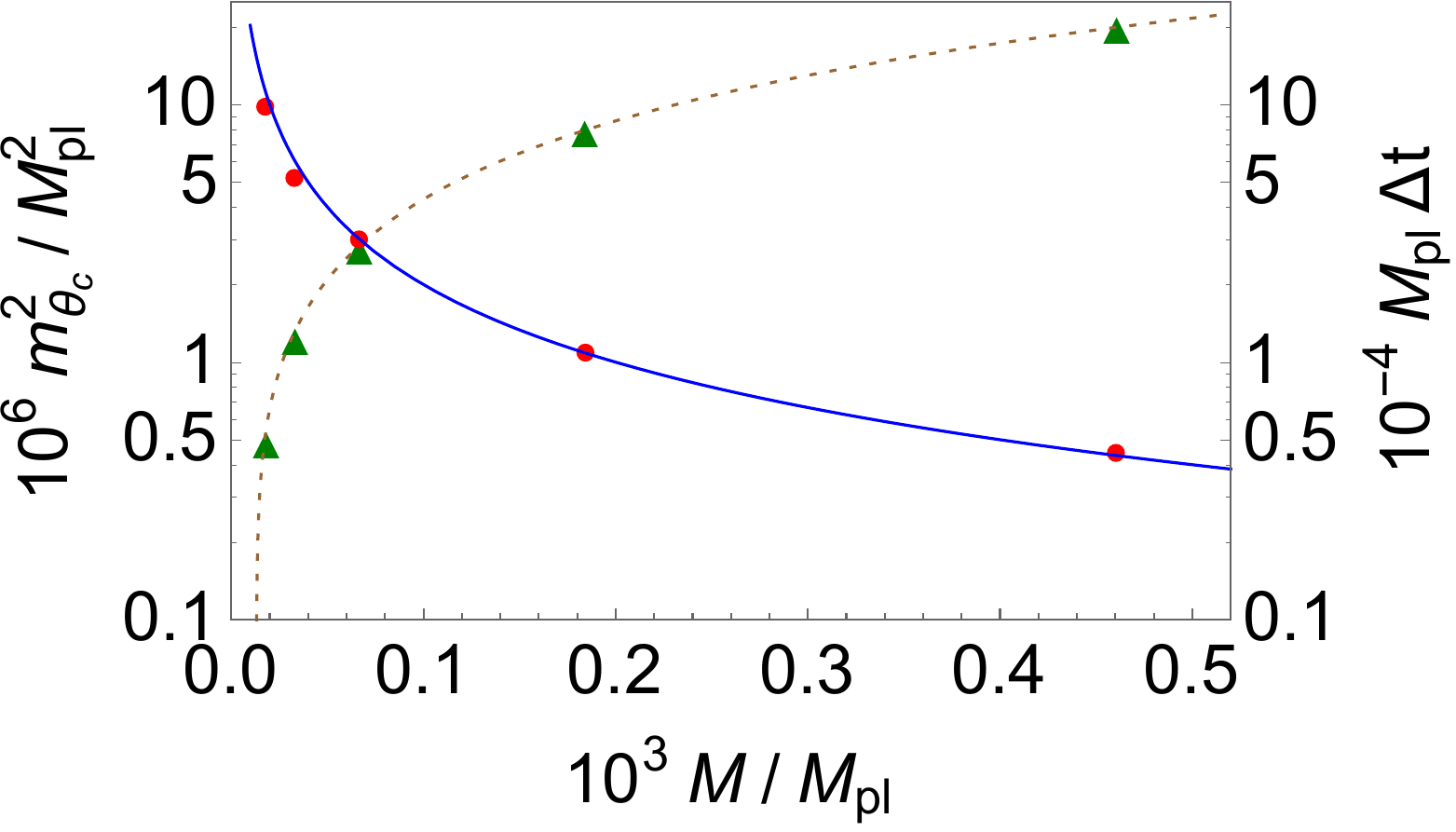}
\caption{
Peak amplitude and timescale of the spike in the effective mass squared of the imaginary field.
The green triangles and the red disks are the numerically obtained peak amplitude and timescale of the mass spike, respectively,
while the brown dashed line and the blue solid line are our analytic estimates (\ref{eq:msp}) and (\ref{eq:dtsp}) with $C_\mathrm{m} \simeq 0.25$ and $C_\mathrm{t} \simeq 2$.
}
\label{fig:Analytic}
\end{figure}

In order to study quantum creation of the NG mode due to the spiky behavior of its mass term, it is convenient to define a canonically normalized scalar field $ \theta_c $ from the phase of the ${\mathcal H}$ field, ${\mathcal H}(x) = h(t)e^{i\theta(x)}/\sqrt{2}$. Since
 the potential $U$ is independent of $\theta$, the relevant part of the Lagrangian reads
\begin{equation}
 \sqrt{-g_\mathrm{E}}{\cal L}_\mathrm{E} \supset 
 - \frac{1}{2} \sqrt{-g_\mathrm{E}} e^{-\alpha \varphi} h^2 g^{\mu\nu}_{\mathrm E} \partial_\mu \theta  \partial_\nu \theta 
 = \frac{1}{2}\dot{\theta}_c^2 - \frac{1}{2a^2}(\nabla\theta_c)^2+\frac{1}{2}\frac{\ddot{F}}{F}\theta_c^2 +\cdots,
\end{equation}
where $\theta_c(x)$ and $F(t)$ are defined as
\begin{align}
 \theta_c(x) \equiv a^{3/2}(t) e^{-\alpha \varphi(t)/2} h(t) \theta(x) \equiv F(t) \theta(x),
\end{align}
 in the Friedmann background, $ds^2 = -dt^2 + a^2(t) d{\vect x}^2$, and a dot denotes differentiation
 with respect to $t$.
 Then we read off the effective mass of the NG mode  as 
\begin{equation}
m_{\theta_c}^2=-\frac{\ddot{F}(t)}{ F(t)}=
-\frac{\alpha}{2} \frac{\partial U}{\partial\varphi} + \frac{e^{\alpha \varphi}}{h}\frac{\partial U}{\partial h} 
-\frac{3}{4}\frac{U}{M^2_\mathrm{pl}} +\frac{5}{24}\frac{1}{M^2_\mathrm{pl}}\left(\dot{\varphi}^2+e^{-\alpha \varphi} \dot{h}^2\right)~, \label{eq:effmass}
\end{equation}
where we have used the background equations~\eqref{eq:eom1}, \eqref{eq:eom2}, and \eqref{eq:eom3}.
While the last two terms are always of the order of the Hubble parameter, the first two terms can be much larger 
when the scalar field trajectory deviates from the valley~\eqref{valley}. 
Figures~\ref{fig:MassEvolution} show the time evolution of $m^2_{\theta_c}$ for our benchmark parameters (A), (B), and (C).
The effective mass gets larger when $\varphi$ gets negative and more or less the spikes still appear even in the case where the $R^2$ term is present.
We can also see that the height of the spikes gets lower and their width gets wider for smaller $M$, 
when the system is more $R^2$-inflation like. 
In Figs.~\ref{fig:Analytic},
we show the heights and the widths of the spikes as a function of $M$ under the observational constraint~\eqref{eq:Mxiconstraint}.
Here we define the width of the spikes as the full width at half maximum. 

The behavior of the spikes can be understood analytically as follows. 
Just before the end of inflation, the energy density of the Universe is dominated by the potential term~\eqref{infpot}. 
Since the potential energy would also dominate the kinetic energy when the $\varphi$ field climbs up the alley $\varphi<0$, 
the potential energy at $h=0$ at the first oscillation can be written as
\begin{align}
U(\varphi, h=0) 
&= C_\mathrm{m}^2 U_\mathrm{inf}, 
\end{align}
which yields with Eq.~\eqref{infpot}
\begin{align}
e^{-\alpha \varphi} 
&=
1 + C_\mathrm{m}\frac{\mtilde}{M} =1 + C_\mathrm{m} \sqrt{\frac{\lambda M_\mathrm{pl}^2}{ 3 \xi^2 M^2 +\lambda M_\mathrm{pl}^2}} 
\quad \text{at} \quad h
= 0. 
\end{align}
Here $C_\mathrm{m} \lesssim 1$ represents the dissipation of the potential energy from the plateau region during inflation to $\varphi$ reaching the largest negative value after a half oscillation.
Then the effective mass of the phase of ${\mathcal H}$ field at $h=0$, which corresponds to the height of the spike, can be estimated as
\begin{align}
\left(m^{\mathrm{sp}}_{\theta_c}\right)^2 
&\simeq 
M^2
\left[
\frac{C_\mathrm{m}}{2} (6 \xi+1) \frac{\mtilde}{M}
- \frac{C_\mathrm{m}^2}{16} \lmk  \frac{\mtilde}{M}\rmk^2
\right].  \label{eq:mspf}
\end{align}
For larger $M (>\sqrt{\lambda} M_\mathrm{pl}/\xi)$, we have $ \left(m^{\mathrm{sp}}_{\theta_c}\right)^2  \simeq C_\mathrm{m} \sqrt{3 \lambda} M M_\mathrm{pl}$. Note that in this expression
for sufficiently large $M \simeq \sqrt{\lambda} M_\mathrm{pl}$ and $\xi \gg 1$\footnote{
These parameters do not respect the perturbativity condition Eq.~\eqref{pertcond}.  } 
for which the single-field approximation gets better, we recover the formula in the pure-Higgs inflation 
$\left(m^{\mathrm{sp}}_{\theta_c}\right)^2 \simeq  \sqrt{3} C_\mathrm{m} \lambda M_\mathrm{pl}^2$~\cite{Ema:2016dny}. 
Taking the observational constraint~\eqref{eq:Mxiconstraint} into account, we obtain 
\begin{align}
\left(m^{\mathrm{sp}}_{\theta_c}\right)^2 
&\approx 
C_\mathrm{m} \sqrt{3 \lambda(M^2 - M_c^2)} M_\mathrm{pl},
\label{eq:msp}
\end{align}
for $\xi \gg 1$.
As discussed above, the duration of the first spike is determined by the period 
when the scalaron stays in the $\varphi < 0$ region,
which is determined by the scalaron mass $M$. 
Therefore, the width of the spikes can also be estimated as
\begin{align}
\Delta t_\mathrm{sp} 
&= 
C_\mathrm{t} M^{-1}.
\label{eq:dtsp}
\end{align}
Again, the formula for the pure-Higgs inflation $\Delta t_\mathrm{sp} \simeq (\sqrt{\lambda} M_\mathrm{pl})^{-1}$
is recovered at $M \simeq \sqrt{\lambda} M_\mathrm{pl}$ when the single field description gets better. 
Note that we cannot use Eqs.~(\ref{eq:mspf}) and (\ref{eq:dtsp}) any more when $M$ significantly exceeds 
the spike timescale inverse in the pure-Higgs case $M \gg \sqrt{\lambda} M_\mathrm{pl}$.
Figure~\ref{fig:Analytic} shows the measured peak amplitude and timescale of the spikes,
as well as our analytic estimates (\ref{eq:msp}) and (\ref{eq:dtsp})
with $C_\mathrm{m} \simeq 0.25$ and $C_\mathrm{t} \simeq 2$.\footnote{
A rough estimate of $C_\mathrm{m}$ goes as follows.
For the pure Higgs or $R^2$ inflation, the potential shape becomes $\propto (1 - e^{-\alpha \varphi})^2$.
The slow roll condition $\max(|\epsilon|,|\eta|) < 1$ breaks down at $e^{-\alpha \varphi} = 2 \sqrt{3} - 3$,
when the inflaton potential energy is $\simeq 0.287$ times its value at the plateau. 
}
The green triangles and the red disks are the values of the amplitude and the timescale estimated from the numerical time evolution, respectively, 
while the brown dashed line and the blue solid line are the predictions of our analytic estimates (\ref{eq:msp}) and (\ref{eq:dtsp}), respectively.
We see that the numerical results coincide with the analytic estimates well.

\subsection{Estimate on particle production}

Let us roughly estimate the energy density of the particles produced in the first spike.
We neglect particle production from the other oscillations because we are mainly interested in the effect of the strongest spike which appears in the pure-Higgs inflation.
A more detailed analysis, including the consequences of other oscillations, will be presented elsewhere~\cite{FutureWork}.

For the estimate of particle production from strong spikes, we can for example refer to the Appendix C of Ref.~\cite{Amin:2015ftc}.
If we describe the strong spike with the  following $\cosh$-type spike function
\begin{align}
m_{\theta_c}^2(t)
&= 
\frac{m}{2\Delta t} \frac{1}{\cosh^2(t/\Delta t)}. 
\end{align}
and the produced field is in the vacuum for $t \to -\infty$, 
its number density after the spike 
is given by
\begin{align}
n_{\theta_c}
&=
\int \frac{d^3k}{(2 \pi)^3}~f_{\theta_c},
\quad
f_{\theta_c}
= 
\cos^2 \left( \frac{\pi}{2} \sqrt{1 + 2m\Delta t} \right) \Big/ \sinh^2(\pi k \Delta t),
\label{eq:fAnalytic}
\end{align}
with $k$ being the wavenumber.
Since we have used the full width at half maximum to estimate $\Delta t_\mathrm{sp}$,
we may identify it as $\Delta t_\mathrm{sp} = \Delta t \times 2\ln(\sqrt{2} + 1)$.
With $\left(m^{\mathrm{sp}}_{\theta_c}\right)^2 \simeq m / 2\Delta t$,
we can estimate the energy density of the produced phase direction as
\begin{align}
\label{eq:energy-density}
\rho_{\theta_c}
&\simeq
\int \frac{d^3k}{(2 \pi)^3}~kf_{\theta_c}
\sim
4.5 \times 10^{-3}
\Delta t_\mathrm{sp}^{-4} \simeq 2.8 \times 10^{-4} \left(\frac{C_\mathrm{t}}{2}\right)^{-4} M^4.
\end{align}
Here we estimated the cosine-squared in the numerator of Eq.~(\ref{eq:fAnalytic}) to be $0.5$.\footnote{
In fact, for $M \gg \tilde{M}$, 
the parameter dependence $m \Delta t \sim \left(m^{\mathrm{sp}}_{\theta_c}\right)^2 \left( \Delta t_\mathrm{sp} \right)^2 
\propto M_\mathrm{pl}/M$
makes the argument of the cosine much larger than unity.}
The estimate (\ref{eq:energy-density}) is in agreement with the general result for particle creation in cosmology obtained in~\cite{Zeldovich:1971mw} 
and with more detailed expression for the rate of particle creation in~\cite{Zeldovich:1977vgo}.
In particular, at the boundary to the strongly-coupled condition $M\simeq \sqrt{4 \pi/\lambda} {\tilde M}$, we have
\begin{align}
\rho_{\theta_c}
&\simeq
4.5 \times 10^2  \left(\frac{\lambda}{0.01}\right)^{-2}\left(\frac{C_\mathrm{t}}{2}\right)^{-4} {\tilde M}^4 \simeq 7.6 \times 10^{-8} \left(\frac{\lambda}{0.01}\right)^{-2}\left(\frac{C_\mathrm{t}}{2}\right)^{-4} {\tilde M}^2 M_{\rm pl}^2. \label{eq:indphrhoex}
\end{align}
which is much smaller than the energy density carried by the inflaton just after the end of inflation is 
$\rho_\mathrm{inf} \simeq U_\mathrm{inf} \simeq  {\tilde M}^2 M_{\rm pl}^2$. 
For smaller $M$, the energy density of the phase direction becomes even smaller. 
Note that this discussion does not rely on the observational condition~\eqref{eq:Mxiconstrainto}. 
Therefore we conclude that, 
even when we add $R^2$ term so that the cut-off scale of the theory becomes the Planck scale, 
the spike still appears and is a real physical phenomenon, 
but the reheating of the Universe does not complete with the violent production of the NG bosons from a single spike.

\section{Discussion and conclusions}
\label{sec:DC}
\setcounter{equation}{0}

We have studied the effective mass of the NG mode for mixed Higgs-$R^2$ model~\cite{Wang:2017fuy,Ema:2017rqn,He:2018gyf,Gundhi:2018wyz} and found that the effective mass has spikes over the preheating process as in the pure-Higgs inflation~\cite{Ema:2016dny}. The set-up is more reliable as the cut-off scale of the model is extended up to $M_\mathrm{pl}$ thanks to the scalaron originated from the $R^2$ term. 
We found that the properties of the spikes are well described by the analytic formula Eqs.~\eqref{eq:msp} and \eqref{eq:dtsp}.  The amplitude of the spikes becomes smaller when the model is more $R^2$-inflation like.
Remarkably,  the energy scale of the spike is well below the cut-off scale of the model, contrary to the case of the pure Higgs inflation,  so
we conclude that the spiky behavior of the NG mode is a real physical phenomenon. 

According to the estimation in  Eq.~\eqref{eq:fAnalytic},
even in the extreme case with the parameters being on the boundary to the strongly-coupled regime, 
the produced energy density of NG boson is much smaller than the total energy density of the Universe.
Thus, the reheating cannot be completed within only one spike (see Eq.~\eqref{eq:indphrhoex}). 
This conclusion is sharply distinctive from the one in the pure-Higgs inflation. 
Although we have worked with the global U(1) scalar ${\cal H}$, we expect that 
our conclusion remains unchanged for the realistic SU(2)$_L \times$ U(1)$_Y$ case. 
Therefore, the parametric resonance of the transverse mode of the gauge bosons~\cite{Bezrukov:2008ut,GarciaBellido:2008ab,Repond:2016sol} 
or the perturbative decay of the scalaron would be the main channel of the depletion of the inflaton quanta. 
Similar analysis can be done in other UV-extension models of the Higgs inflation. 
The detailed study on the (p)reheating process will be presented elsewhere~\cite{FutureWork}.

\section*{Acknowledgments}
We thank Y.-F. Cai, S. Pi, Y. Watanabe, and Y.-P. Wu for useful discussion. 
MH was supported by the Global Science Graduate Course (GSGC) program of the University of Tokyo. 
RJ and KK were supported by IBS under the project code, IBS-R018-D1. 
KK thanks IBS-CTPU for kind hospitality during the completion of this work. The work of SCP was supported in part by the National Research Foundation of Korea (NRF) grant funded by the Korean government (MSIP) (No.2016R1A2B2016112) and (NRF-2018R1A4A1025334).
AS acknowledges RESCEU hospitality as a visiting professor. He was also partially supported by the RFBR grant 17-02-01008. The work of JY was supported by JSPS KAKENHI, Grant-in-Aid for Scientific Research 15H02082 and Grant-in-Aid for Scientific Research on Innovative Areas 15H05888.  

\small
\bibliography{ref}

\end{document}